\documentclass[twocolumn, superscriptaddress, prb, showpacs]{revtex4}

\usepackage{graphicx}
\usepackage{amsmath}

\def\A{{\bf A}}
\def\a{{\bf a}}
\def\b{{\bf b}}
\def\G{{\bf G}}
\def\I{{\bf I}}
\def\k{{\bf k}}
\def\R{{\bf R}}
\def\r{{\bf r}}
\def\x{{\bf x}}
\def\0{{\bf 0}}
\def\Dcr{\ensuremath{(\Delta/t_2)_{\text{cr}}}}
\def\ket#1{\vert#1\rangle}
\def\bra#1{\langle#1\vert}
\def\me#1#2#3{\langle#1\vert#2\vert#3\rangle}
\def\ip#1#2{\langle#1\vert#2\rangle}

\def\intk{\int_{\rm BZ} \!\!\!\! d{\bf k} \;}
\def\park{\partial_{\bf k}}
\def\nn{\nonumber\\}

\begin{document}

\title{The insulator/Chern-insulator transition in the Haldane model}
\author{T. Thonhauser}
\affiliation{Department of Physics and Astronomy, Rutgers, The State 
University of New Jersey, Piscataway, New Jersey 08854, USA.}
\author{David Vanderbilt}
\affiliation{Department of Physics and Astronomy, Rutgers, The State 
University of New Jersey, Piscataway, New Jersey 08854, USA.}
\date{\today}

\begin{abstract}
We study the behavior of several physical properties of the Haldane
model as the system undergoes its transition from the normal-insulator
to the Chern-insulator phase.  We find that the density matrix has
exponential decay in both insulating phases, while having a power-law
decay, more characteristic of a metallic system, precisely at the phase
boundary.  The total spread of the maximally-localized Wannier functions
is found to diverge in the Chern-insulator phase. However, its
gauge-invariant part, related to the localization length of Resta and
Sorella, is finite in both insulating phases and diverges as the phase
boundary is approached.  We also clarify how the usual algorithms for
constructing Wannier functions break down as one crosses into the
Chern-insulator region of the phase diagram.
\end{abstract}

\maketitle

\section{Introduction}

The band structure of any insulator is characterized by a certain
discrete topological index known as the Chern invariant\cite{Thouless98}
which encodes information about the phase evolution of the Bloch
functions around the boundary of the Brillouin zone (see
Sec.~\ref{sec:chern}).  Insulators can thus be classified as ``normal
insulators'' or ``Chern insulators'' depending on whether or not the
Chern invariant vanishes.  The latter case requires breaking of
time-reversal symmetry, so insulating ferromagnets and ferrimagnets
could be candidates for Chern insulators.  While models for Chern
insulators can be constructed theoretically, \cite{Haldane88} no
experimental realizations are yet known to occur in nature.  A Chern
insulator, if found to exist, would have the remarkable feature of
showing a quantum Hall effect in the \emph{absence} of a macroscopic
magnetic field.  Hence, Chern insulators may also be referred to as
``quantum Hall insulators.''

Although the basic theory of Chern insulators was formulated in the
1980's, not much is known theoretically about the general features of
the electronic band structure of such insulators. In the last 15 years or
so, the theory of normal insulators has been greatly enriched by a
deeper understanding of electric polarization,\cite{KSV} orbital
magnetization,\cite{Thonhauser05,Ceresoli06} linear-scaling theory,
methods for constructing Wannier functions,\cite{Marzari97} the spatial
decay of Wannier functions and of the one-particle density
matrix,\cite{He01} and related measures of
localization.\cite{Resta99,Souza00,Resta02} However, all of this work
implicitly assumed the presence of time-reversal symmetry, and thus was
limited to the case of normal insulators.  It is therefore of
considerable interest to revisit many of these same issues, and to
reconsider whether, or how, the previous conclusions generalize to the
case of Chern insulators.  For example, what are the decay properties of
the one-particle density matrix in a Chern insulator?  Can Wannier
functions be constructed, and if not, in what way do the usual
construction procedures fail?  If one inspects closely related measures
of localization such as the gauge-invariant part of the Wannier spread
functional,\cite{Marzari97} the localization length of Resta and
Sorella,\cite{Resta99}  or the second-cumulant moment of the electron
distribution, \cite{Souza00, Resta02} does the localization remain
finite in a Chern insulator, or does it diverge?

Furthermore, an intriguing feature of Chern-insulating systems is that a
phase boundary separating the Chern insulator from a normal insulator
may occur.  Such a normal-insulator/Chern-insulator (NI/CI) transition
is an example of a class of topological transitions that have become of
considerable current interest, at which a topological invariant changes
discontinuously across a phase boundary.\cite{Hatsugai05} Such
transitions normally appear within the theory of correlated
states.\cite{Wen95, Wen00} The NI/CI transition, on the other hand,
occurs in a non-interacting context, and can therefore be studied at a
level of detail, and tested with numerical calculations, in a way that
is difficult for correlated models.  In addition, Chern insulators are
closely related to so-called ``spin Hall insulators,'' which have also
been the subject of a recent surge of interest.\cite{PhysicsToday} Thus,
there is a special opportunity associated with the study of this
particular topological insulator/insulator transition.

In this paper, we investigate several aspects of the electronic
structure near the NI/CI transition in the two-dimensional Haldane
model.\cite{Haldane88} We choose the Haldane model because it is one of
the simplest models that exhibits a quantum-Hall insulator state. The
underlying idea of this model is to break time-reversal symmetry so that
the transverse conductivity  $\sigma_{xy}$, which is odd under
time-reversal, can become nonzero. Usually, the quantum Hall effect is
associated with a gap at the Fermi level resulting from a splitting of
the spectrum into Landau levels by a macroscopic magnetic field. In the 
Haldane model, however, there is a degeneracy between the valence and
conduction bands at certain high-symmetry $\k$-points when both
inversion and time-reversal symmetry are present.  If a gap is opened by
the breaking of inversion symmetry, the system becomes a normal
insulator. However, if the gap opens as a result of breaking
time-reversal symmetry, the system turns into a Chern insulator.

We have organized this paper as follows. In Sec.~\ref{sec:chern} we
introduce the Chern invariant, which will henceforth be used to classify
the state of our system. The basics of the Haldane model are reviewed in
Sec.~\ref{sec:haldane}. Thereafter, we focus on the  problems occurring 
when constructing Wannier functions for Chern insulators
(Sec.~\ref{sec:WF}), the behavior of the spread functional
(Sec.~\ref{sec:spread}), and the decay of the density matrix
(Sec.~\ref{sec:densmat}) as the system transitions from the normal
insulator phase into the Chern-insulator phase. We conclude and give an
outlook in Sec.~\ref{sec:conclusions}.

\section{The Chern invariant}
\label{sec:chern}

We restrict ourselves to the case of a one-particle Hamiltonian $H$
having Bloch eigenvalues $\epsilon_{n\k}$ and eigenstates
$\ket{\psi_{n\k}}$.  The cell-periodic part of the Bloch function 
$u_{n\k}(\r)=e^{-i\k\cdot\r}\psi_{n\k}(\r)$ is then an eigenfunction of
the effective Hamiltonian $H(\k)=e^{-i\k\cdot\r} H e^{i\k\cdot\r}$. We
consider electrons to be spinless, but factors of two can easily be
inserted for non-interacting spin channels.

We can now define the Chern invariant\cite{Thouless98} for an insulator,
defined here as a system with a gap in the single-particle density of
states separating occupied and unoccupied states, to be
\begin{equation}\label{equ:chern}
{\bf C} = \frac{i}{2\pi} \intk \sum_n^\text{occ}
\me{\park u_{n\k}}{\times}{\park u_{n\k}}\;,
\end{equation}
where BZ denotes an integral over the Brillouin zone and $\park =
\partial/\partial \k$. The cross product notation in
Eq.~(\ref{equ:chern}) implies, for example, that $C_z$ contains terms
involving $\ip{\partial_{k_x} u_{n\k}}{\partial_{k_y} u_{n\k}}
-\ip{\partial_{k_y} u_{n\k}}{\partial_{k_x} u_{n\k}}$. For
non-interacting electrons,\cite{Explan-ni,Thouless82} the Chern
invariant is quantized in units of reciprocal-lattice vectors $\G$. For
the case of a two-dimensional system with only a single occupied band,
Eq.~(\ref{equ:chern}) becomes
\begin{equation}\label{equ:chern2D}
C = \frac{i}{2\pi} \intk\me{\park u_{\k}}{\times}{\park u_{\k}}\;.
\end{equation}
In two dimensions the Chern invariant is a pseudo-scalar called the
\emph{Chern number} which can only take integer values. Alternatively,
we can write the Chern number in terms of the Berry connection
$\A(\k)=i\me{u_{\k}}{\park}{u_{\k}}$ and the Berry curvature
$\Omega(\k)=\nabla_\k\times\A(\k)$ as
\begin{equation}\label{equ:chern_omega}
C = \frac{1}{2\pi} \intk\Omega(\k) =
\frac{1}{2\pi}\oint_{\rm BZ} \!\!\!\! d\k\cdot\A(\k)\;.
\end{equation}
A Chern insulator is now simply defined as an insulator with a nonzero
Chern invariant. Conversely, we define a normal insulator to be an
insulator with zero Chern invariant. Hence, the NI/CI transition is
characterized by a change of the Chern invariant from zero to a nonzero
value.

The Chern invariant of Eqs.~(\ref{equ:chern}) and (\ref{equ:chern2D}) is
gauge invariant,\cite{Ceresoli06} i.e., invariant with respect to the
choice of phases of the $|u_{n\k}\rangle$, or in the more general
multiband case, the choice of unitary rotations applied to transform the
occupied states among themselves at a given $\k$. It can be shown that
in normal insulators it is always possible to make a gauge choice such
that the Bloch orbitals are periodic in $\k$-space (i.e.,
$\ket{\psi_{n\k\!+\!\G}} = \ket{\psi_{n\k}}$) and smooth in $\k$ (i.e.,
continuous and differentiable), whereas no such choice is possible for a
Chern insulator.\cite{Panati06}

\section{The Haldane model}
\label{sec:haldane}

\begin{figure}
\begin{center}
{\bf (a)} \includegraphics[width=4cm]{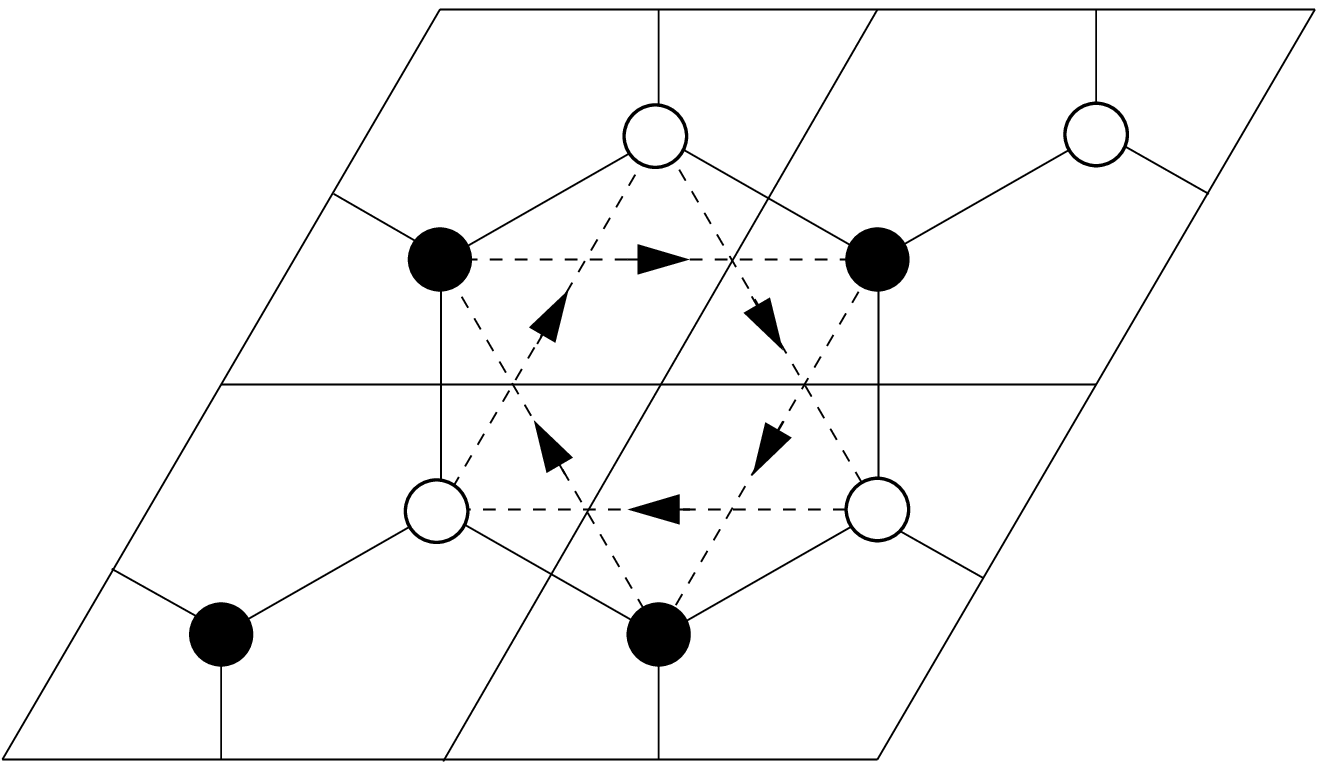}\quad
{\bf (b)} \includegraphics[width=2.7cm]{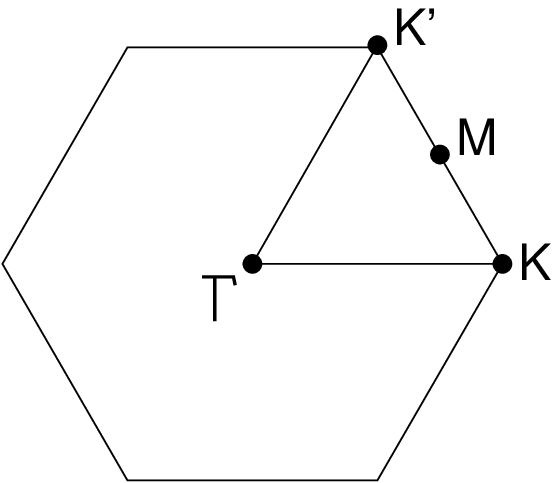}
\end{center}
\caption{\label{fig:model} (a) Four unit cells of the Haldane model.
Filled (open) circles have site energy $-\Delta$ ($+\Delta$). The
first-nearest neighbor hopping $t_1$ is real, while the second-nearest
neighbor hopping $t_2 e^{i\varphi}$ has a complex phase. Arrows indicate
the direction of positive phase hopping. The Wigner-Seitz unit cell
corresponds to the hexagon in the center of the plot. (b) First
Brillouin zone of the Haldane model with high-symmetry points marked.}
\end{figure}

Here we provide a brief review of Haldane's model and its properties, as
discussed in detail in Ref.~[\onlinecite{Haldane88}]. As illustrated in
Fig.~\ref{fig:model}, the Haldane model is comprised of a honeycomb
lattice having two tight-binding sites per cell with site energies $\pm
\Delta$, a real first-neighbor hopping $t_1$, and a complex
second-neighbor hopping $t_2e^{\pm i\varphi}$. The model can also be
thought of as consisting of two sublattices $A$ and $B$ corresponding to
the sites with energies $+\Delta$ and $-\Delta$, respectively. Note that
the \emph{macroscopic} magnetic flux through the unit cell is indeed
zero, resulting in a vanishing macroscopic magnetic field. This follows
directly from the fact that the first-nearest neighbor hopping is real
and no phase is picked up when hopping around the Wigner-Seitz unit
cell. This, however, does not rule out a \emph{microscopic} magnetic
field that averages to zero over the unit cell. Note that the wavevector
$\k$ is still a good quantum number under these conditions.

Let $\a_1, \a_2$, and $\a_3$ be the vectors pointing from a site of the
$B$ sublattice to its three nearest $A$ neighbors, such that  $\hat{\bf
z}\cdot\a_1\times\a_2>0$ and $\hat{\bf x}\cdot\a_1>0$. If we furthermore
define the vectors $\b_1=\a_2-\a_3$, $\b_2=\a_3-\a_1$, and
$\b_3=\a_1-\a_2$, then the Hamiltonian of the Haldane model can be
written as
\begin{eqnarray}\label{equ:hamiltonian}
\lefteqn{H(\k) = \I\;2t_2\cos\varphi\sum_i\cos(\k\cdot\b_i)}\nn
&&{}+\boldsymbol{\sigma}_1\;t_1\sum_i\cos(\k\cdot\a_i) +
  \boldsymbol{\sigma}_2\;t_1\sum_i\sin(\k\cdot\a_i)\nn
&&{}+\boldsymbol{\sigma}_3\Big(\Delta-2t_2\sin\varphi\sum_i\sin(\k\cdot\b_i)\Big)\;,
\end{eqnarray}
where the $\boldsymbol{\sigma}_i$ are the Pauli matrices and $\I$ is the
identity matrix.

\begin{figure}
\begin{center}
\includegraphics[width=\columnwidth]{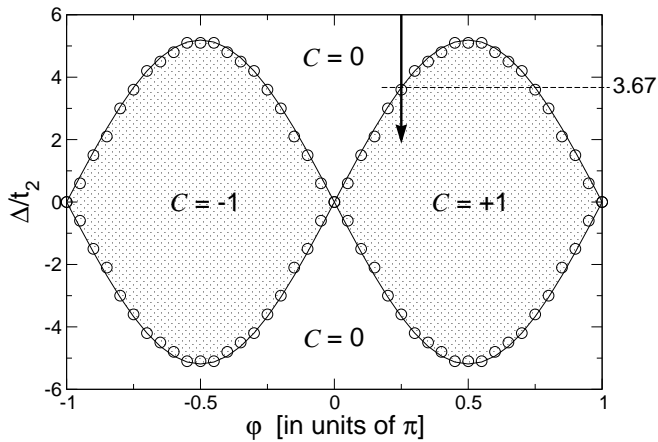}
\end{center}
\caption{\label{fig:phase} Chern number of the bottom band of the
Haldane model as a function of the parameters $\varphi$ and $\Delta/t_2$
($t_1=1$, $t_2=1/3$). Analytic results are plotted as solid sinusoidal
lines, whereas numerical results are depicted by circles. The vertical
line shows the range of parameters that we have chosen for all our
calculations.}
\end{figure}

The Chern number can now be calculated analytically or numerically
according to Eq.~(\ref{equ:chern2D}) or (\ref{equ:chern_omega}). For our
tests, we have chosen a lattice constant equal to unity, $t_1=1$, and
$t_2=1/3$. If the Chern number of the bottom band is mapped out as a
function of the remaining model parameters $\varphi$ and $\Delta/t_2$,
we obtain the Haldane phase diagram shown in Fig.~\ref{fig:phase}.
Since we are interested in studying the transition from a normal
insulator to a Chern insulator, we choose for all our calculations below
a path in the phase diagram that crosses the phase boundary.
Specifically, we traverse the vertical line in Fig.~\ref{fig:phase}
where the phase $\varphi$ is fixed at $\pi/4$ and $\Delta/t_2$ is
reduced from 6 to 2. At the critical value
$\Dcr=3\sqrt{3}\sin(\pi/4)\approx 3.67$, the phase boundary is crossed.

\begin{figure}
\begin{center}
\includegraphics[width=\columnwidth]{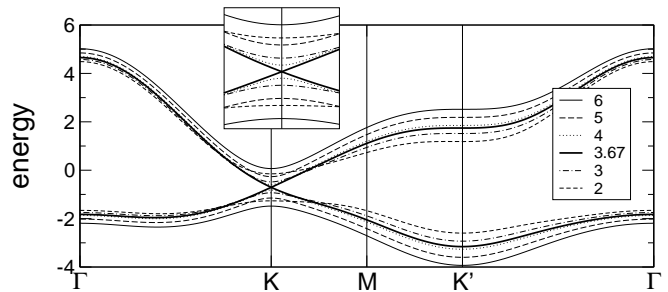}
\end{center}
\caption{\label{fig:bs} Band structure of the Haldane model along some
high-symmetry lines for several values of $\Delta/t_2$ along the path
marked in Fig~\ref{fig:phase}. The inset shows a magnification of the
bands at $K$. Note that at \Dcr\ the dispersion is linear.}
\end{figure}

The band structure of the Haldane model is plotted in Fig.~\ref{fig:bs}
along some high-symmetry lines in the Brillouin zone (see
Fig.~\ref{fig:model}b). It shows a remarkable feature as the system
passes through \Dcr. In the normal-insulator region, the two bands are
separated by a finite gap. As the critical value is approached, the gap
at $K$ gets smaller and smaller. Finally, exactly at \Dcr\ the bands
touch at $K$ in such a way that the dispersion relation is linear. Such
points are also referred to as \emph{Dirac} points. When going further
into the Chern-insulator region, the bands separate again.  Note that
our specific choice of $t_1=1$ and $t_2=1/3$ prevents the bands from
overlapping. If $\Delta$ and $t_2\sin\varphi$ are both chosen to be
zero, two Dirac points form at $K$ and $K'$, and the Haldane model then
becomes an appropriate model for a graphene sheet.\cite{Wilson06}

In the normal-insulator region of the Haldane model the Chern number of
each band is zero, so that the total Chern number (the sum of the Chern
numbers of the upper and lower bands) is obviously also zero.  When the
phase boundary is crossed, the Chern numbers of the upper and lower
bands become $\pm 1$, but their sum still remains zero. The closure and
reopening of the gap as the NI/CI boundary is crossed corresponds to the
``donation'' of a Chern unit from one band to another through the
temporarily formed Dirac point.  In the present case, the total Chern
number must always remain zero because the model, having a tight-binding
form, assumes Wannier representability of the overall band space, and a
non-zero Chern number is inconsistent with such an assumption. More
generally, the total Chern number of a group of bands should not change
when a gap closure and reopening occurs among the bands of the group, as
long as the gaps between this group and any lower or higher bands
remains open.

\begin{figure}
\begin{center}
\includegraphics[width=\columnwidth]{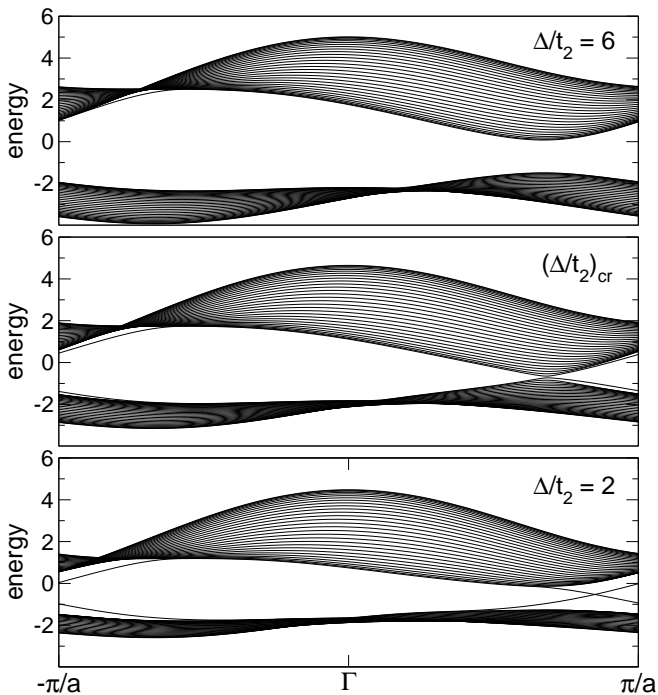}
\end{center}
\caption{\label{fig:surface_bs} Energy vs.\ wavevector $k_y$ for the
Haldane model in a strip geometry 30 cells wide along $\b_3$ direction
and extending infinitely along $\b_2$ direction. For $\Delta/t_2<\Dcr$
(bottom panel), chiral edge states are visible.}
\end{figure}

It is possible to argue on very general grounds that a finite sample cut
from a Chern insulator must have conductive channels, otherwise known as
chiral edge states, that circulate around the perimeter of the
sample\cite{Haldane06} in much the same way as for the quantum Hall
effect.\cite{Halperin82,Yosh} It is therefore of interest to investigate
the electronic structure of the Haldane model from the point of view of
the surface band structure. We consider a sample that is finite in the
$\b_3$ direction (specifically, 30 cells wide) and has periodic boundary
conditions along the $\b_2$ direction (the $\b_i$ are defined above
Eq.~(\ref{equ:hamiltonian})); its states can be labeled by a wavevector
$k_y$ running from $-\pi/a$ to $+\pi/a$, where $a$ is the repeat unit in
the $y$ direction.  The energy eigenvalues are plotted vs.\ $k_y$ for
several values of $\Delta/t_2$ in Fig.~\ref{fig:surface_bs}. At first
sight, the surface band structure shows qualitatively the same
information as the bulk band structure in Fig.~\ref{fig:bs}. For
$\Delta/t_2=6$, the  valence and conduction bands are separated by a
finite gap. At the Chern transition a Dirac point forms, showing the
characteristic linear dispersion expected around such a point.  However,
when we go deeper into the Chern insulator, the surface band structure
reveals a new behavior: one surface band now crosses from the lower
manifold to the upper one with increasing $k_y$, and another crosses in
the opposite direction. Further inspection shows that the upgoing and
downgoing states are localized to the right and left surfaces of the
strip, respectively.  Thus, if the Fermi level lies in the bulk gap,
there will be metallic states with Fermi velocities parallel to the
surfaces and with opposite orientation, i.e., a chiral
(counterclockwise) circulation of edge states around the perimeter of
the sample, as expected.

\section{Breakdown of Wannier-function construction at the Chern
transition}\label{sec:WF}

We now study aspects of the NI/CI transition that are related to Wannier
functions (WFs) and electron localization. We expect that in the
normal-insulator phase, it should be straightforward to construct
Wannier functions via a $\k$-space construction.  The term ``Wannier
function'' is usually applied only in the case of periodic systems, but
for finite samples one can construct well localized Boys
orbitals\cite{Boys60} which play the same role and which map onto the
WFs in the thermodynamic limit $n\to\infty$.  Thus, if we cut a finite
sample from a normal-insulator realization of the Haldane model, we also
expect it to be straightforward to construct such Boys orbitals.  The
question then arises as to what, precisely, will ``go wrong'' with these
procedures if one tries to do the same on the Chern-insulator side of
the transition. In particular, for a finite sample cut from the Haldane
model, it is unclear how the system would ``know'' whether the finite
sample corresponds to the normal-insulator or Chern-insulator side of
the transition, and how the construction would break down in the latter
case.  In this Section, we investigate these issues, first in the
context of the real-space construction, and then later from the
$\k$-space point of view.

We start, then, by considering finite $n\times n$ samples of the Haldane
model. We can interpret Fig.~\ref{fig:model} as showing a picture of a
finite sample of size $n=2$; we study similarly-constructed samples of
size $n=$10, 20, 30, and 40.  For each sample, the Boys orbitals are
constructed as follows. We define the projection operator onto the
occupied states as
\begin{equation}
P=\sum_n^{\text{occ}}\ket{\psi_n}\bra{\psi_n}\;,
\end{equation}
and we choose a set of well-localized ``trial'' orbitals
$\ket{t_\alpha}$, equal in number to the number of occupied states, that
we want the Boys orbitals to be roughly modeled after.  We then
construct the projected trial functions $\ket{y_\alpha} =
P\ket{t_\alpha}$.  Since $\rho(\r,\r')=\langle\r|P| \r'\rangle$ is
expected to decay exponentially in $|\r-\r'|$ for an insulator (see
Sec.~\ref{sec:densmat}), we expect the $\ket{y_\alpha}$ to be localized
as well, and as long as they are not overcomplete they will span the
occupied space of interest. However, they are not orthonormal, so the
last step is to carry out a symmetric
orthonormalization.\cite{Carlson57} This is done by computing the
overlap matrix
\begin{equation}
S_{\alpha\beta} = \bra{y_\alpha}y_\beta\rangle
\end{equation}
and then constructing the final Boys orbitals $\ket{\omega_\alpha}$ as
\begin{equation}\label{equ:boys}
\ket{w_\alpha}=\sum_\beta (S^{-1/2})_{\beta\alpha}\,\ket{y_\beta}\;.
\end{equation}

\begin{figure}
\begin{center}
\includegraphics[width=\columnwidth]{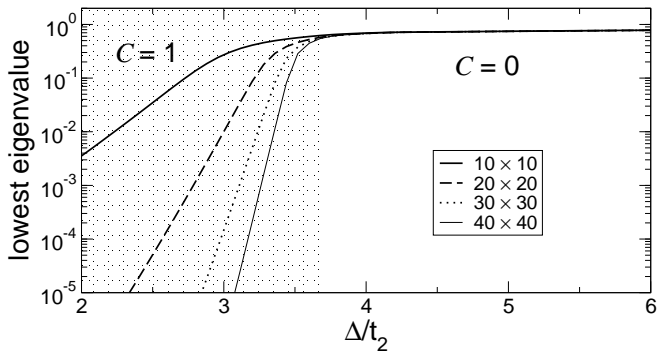}
\end{center}
\caption{\label{fig:lowest} Lowest eigenvalue of the overlap matrix
$S_{\alpha\beta}=\bra{y_\alpha}y_\beta\rangle$ as a function of the
parameter $\Delta/t_2$ for different sample sizes $n\times n$.  Note
the logarithmic scale.}
\end{figure}

In the context of the Haldane model, it is natural to choose the trial
functions to be a set of $\delta$-functions located on the sites of the
lower-energy sublattice.  With this choice, we can now study the lowest
and highest eigenvalue of $S_{\alpha\beta}$ as the parameter
$\Delta/t_2$ traverses the path shown in Fig.~\ref{fig:phase}. While the
highest eigenvalue remains very close to 1, the lowest eigenvalue drops
and rapidly approaches zero in the Chern-insulator region, i.e., for
$\Delta/t_2$ values below the critical value of $\Dcr\approx 3.67$, as
shown in Fig.~\ref{fig:lowest}. The slope of the ``drop''  depends on
the size of the sample and becomes steeper as the sample size gets
larger.  For any given value of $\Delta/t_2<\Dcr$, the lowest eigenvalue
appears to approach zero exponentially with sample size. When the
eigenvalue becomes too small, the inversion to obtain $S^{-1/2}$ becomes
ill-conditioned, and the symmetric orthonormalization in
Eq.~(\ref{equ:boys}) can no longer be carried out. It follows that Boys
orbitals cannot be constructed in the Chern-insulator phase, at least
not using this approach.

We now change perspective and look at the problem from the $\k$-space
point of view, where we find that something similar happens. WFs for
periodic samples are defined by
\begin{equation}\label{equ:WFdef}
\ket{\R n} = \frac{\Omega}{(2\pi)^3}
\intk e^{-i\k \cdot \R} \ket{\psi_{n\k}}\;,
\end{equation}
where the inverse relation is
\begin{eqnarray}
\ket{\psi_{n\k}} = \sum_\R e^{i\k \cdot\R} \ket{\R n}\;.
\end{eqnarray}
In this notation $\ket{\R n}$ refers to the $n$'th WF in cell $\R$.

As mentioned previously, for systems with zero Chern invariant, the 
Bloch orbitals can always be chosen to obey a smooth and periodic gauge
$\ket{\psi_{n\k\!+\!\G}} = \ket{\psi_{n\k}}$. However, if the Chern
invariant becomes nonzero, this choice is no longer
possible.\cite{Panati06} In this case it is possible to make a periodic
gauge choice that is smooth {\it almost} everywhere, but there must be
singularities (``vortices'') somewhere in the interior of the BZ.  For
example, in two dimensions, assume a gauge choice that is periodic and
also smoothly defined everywhere in the BZ except in a small disk
located somewhere in the interior of the BZ.  The periodic gauge choice
implies that $\oint d\k\cdot\A(\k)$ around the perimeter of the BZ must
vanish.  Applying Stoke's theorem as in Eq.~(\ref{equ:chern_omega}), but
now to the region {\it excluding} the small disk, implies that $\oint
d\k\cdot\A(\k)$ around the circumference of the small disk must approach
$-C$ in the limit that the disk becomes small. For the Chern phase
($C\ne0$), this implies that there must be a vortex singularity in the
phase choice inside the disk.  If one attempts to construct WFs naively
using Eq.~(\ref{equ:WFdef}), one then finds that the discontinuity in
the phase choice of $|u_{n\k}\rangle$ at the vortex in $\k$-space leads
to the destruction of exponential localization of the WFs in real space.

We have investigated how this problem manifests itself if one attempts
to construct WFs using standard $\k$-space methods.  Similar to the
approach described in Ref.~[\onlinecite{Marzari97}], we again adopt a
projection method in which one chooses trial Bloch-like functions
$\ket{t_\k}$ that are smooth and periodic in $\k$-space. This can be
done by constructing the $\ket{t_\k}$ from a set of real-space trial
functions $\ket{t_\alpha}$, i.e., $t_\k(\r)=\sum_\R
e^{i\k\cdot\R}\,t_\alpha(\r-\R)$.  Then one can construct projected
states $\ket{y_\k}$ via
\begin{equation}
\ket{y_\k} = P\ket{t_\k} =
\sum_{\k'}\ket{\psi_{\k'}}\bra{\psi_{\k'}}t_\k\rangle
= \ket{\psi_{\k}}\bra{\psi_{\k}}t_\k\rangle
\label{equ:ykeq}
\end{equation}
and orthonormalized projected states
\begin{equation}
\ket{w_\k} = s(\k)^{-1/2}\,\ket{y_\k}\;,
\label{equ:wkeq}
\end{equation}
where
\begin{equation}
s(\k)=\bra{y_\k}y_{\k}\rangle=|\bra{t_\k}y_{\k}\rangle|^2\;.
\label{equ:skeq}
\end{equation}
The WFs are then constructed by Fourier transforming to real space using
Eq.~(\ref{equ:WFdef}) with $\ket{w_\k}$ substituted for $\ket{\psi_\k}$.
Clearly, if $s(\k)$ should vanish at some $\k$, this procedure would
fail.

\begin{figure}
\includegraphics[width=\columnwidth]{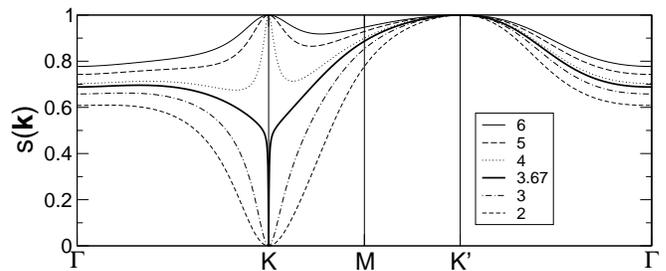}
\caption{\label{fig:s_k}Plot of $s(\k)$ of Eq.~(\protect\ref{equ:skeq})
along some high-symmetry lines for
several values of $\Delta/t_2$.}
\end{figure}

We can now study what happens if this construction procedure is applied
to the Haldane model.  We again use trial functions that are
$\delta$-functions located on the lower-energy sites. We study the
behavior of $s(\k)$ as a function of $\k$ throughout the BZ, while
varying $\Delta/t_2$ along the line in Fig.~\ref{fig:phase}. Results for
$s(\k)$ are plotted along some high-symmetry lines in 
Fig.~\ref{fig:s_k}. In the normal-insulator region, we find
$0<s(\k)\le1$ for all $\k$. After the phase boundary has been crossed at
\Dcr, we find that there is one point $\k_a$ in the BZ for which
$s(\k_a)=0$.  There is also one point $\k_b$ for which $s(\k_b)=1$
exactly.  In our numerical calculations, the locations of $\k_a$ and
$\k_b$ coincide with the points $K$ and $K'$ respectively. By
experimenting with different trial functions, we have found that the
precise locations of the minimum and maximum may deviate from $K$ and
$K'$, and the value at the maximum may be less than unity. However, we
always find a point $\k_a$ at which $s(\k_a)=0$.  This is the point at
which $\ip{\psi_\k}{t_\k}=0$; the robustness of such a zero-crossing can
be understood heuristically by realizing that by adjusting the two
parameters $k_x$ and $k_y$, the real and imaginary parts of the complex
scalar $\ip{\psi_\k}{t_\k}$ can generically both be made to vanish. 
{From} Eqs.~(\ref{equ:ykeq}--\ref{equ:skeq}) it follows that the phase
of $\ket{y_\k}$ evolves by $2\pi$ as one circles around $\k_a$, so that
a vortex-like singularity is generated in the phase of $\ket{w_\k}$
about $\k_a$, with $\ket{w_\k}$ becoming ill-defined precisely at
$\k_a$.  Thus, the construction of well-localized WFs is no longer
possible.

\begin{figure}
\begin{center}
\includegraphics[width=\columnwidth]{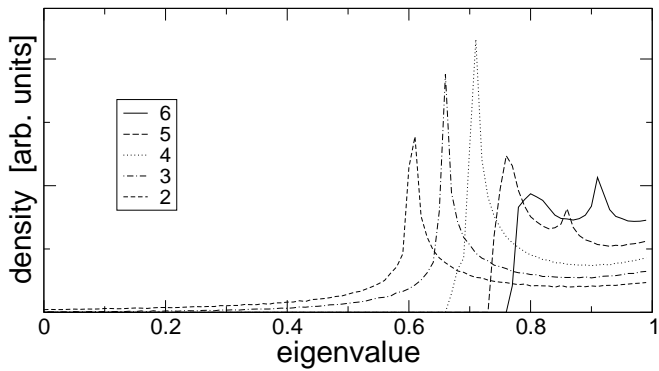}
\end{center}
\caption{\label{fig:dos} Density of $s(\k)$ values as a function of
$\Delta/t_2$ obtained for periodic samples with a $300\times 300$
$\k$-mesh.}
\end{figure}

Instead of focusing only on the lowest eigenvalue, we plot in
Fig.~\ref{fig:dos} the ``density of overlap values'' $s(\k)$. In the
normal-insulator region of $\Delta/t_2$, one sees typical 2D van Hove
singularities, and in particular, a well-defined minimum above zero.  In
the Chern-insulator region of $\Delta/t_2$, on the other hand, the
density of overlap values shows a tail extending all the way to zero.

In summary, when the system is in its normal-insulator phase, the
construction of Boys orbitals for finite samples, or of WFs for periodic
samples, can be carried out in the usual way using a projection method. 
However, once the NI/CI phase boundary has been crossed, such a
construction is bound to fail because of singularities that appear in
the overlap matrices in both the real-space finite-sample and $\k$-space
extended-sample approaches.

\section{The spread functional}
\label{sec:spread}

Another quantity that shows interesting behavior as the phase boundary
is crossed is the spread functional $\Omega$ in real space, defined by
Marzari and Vanderbilt (MV)\cite{Marzari97} to be
\begin{equation}
\Omega =
\sum_n\Big[\me{\0n}{r^2}{\0n}-\me{\0n}{\r}{\0n}^2\Big]\;,
\end{equation}
where $\ket{\0n}$ refers to the WF $\ket{\R n}$ for band $n$ in the home
unit cell $\R=\0$ and the sum is over occupied bands of the insulator.
The spread functional is a measure of how ``spread out'' or delocalized
the WFs are.  In the remainder of this section, we specialize for
simplicity to the case of a single band in two dimensions, so that
$\Omega=\me{\0}{r^2}{\0}-\me{\0}{\r}{\0}^2$. MV showed that the spread
functional can be decomposed as $\Omega=\Omega_I+\widetilde{\Omega}$,
where
\begin{equation}
\Omega_I=\me{\0}{r^2}{\0}-\sum_\R |\me{\0}{\r}{\R}|^2
\label{equ:ir}
\end{equation}
and
\begin{equation}
\widetilde{\Omega}=\sum_{\R\ne\0}|\me{\0}{\r}{\R}|^2
\label{equ:dr}
\end{equation}
are gauge-invariant and gauge-dependent contributions, respectively. The
gauge-invariant part has been shown to be a useful measure for
characterizing the system: $\Omega_I$ is finite in insulators and
diverges in metals.\cite{Resta99}

MV also gave corresponding $\k$-space expressions for the two parts of
the functional. Defining the metric tensor $g_{\mu\nu}= {\rm
Re}\,\me{\partial_\mu u_\k}{Q_\k}{\partial_\nu u_\k}$ where
$Q_\k=1-\ket{u_\k}\bra{u_\k}$ (and $\partial_\mu=\partial/\partial
k_\mu$), these two quantities can be rewritten as
\begin{equation}
\Omega_I=\frac{\cal A}{(2\pi)^2}\intk{\rm Tr}[\,g(\k)\,]
\label{equ:ik}
\end{equation}
and
\begin{equation}
\widetilde{\Omega}=\frac{\cal A}{(2\pi)^2}\intk|\A(\k)-\bar{\A}|^2\;,
\label{equ:dk}
\end{equation}
where $\cal A$ is the unit cell area, ${\rm Tr}\,[g]=g_{xx}+g_{yy}$, and
$\bar{\A}$ is the BZ average of $\A(\k)$ defined just above
Eq.~(\ref{equ:chern_omega}).

In the case of a Chern insulator, the use of the real-space expressions
(\ref{equ:ir}--\ref{equ:dr}) becomes problematic, since well-localized
WFs cannot be constructed.  Nevertheless, the reciprocal-space
expressions (\ref{equ:ik}--\ref{equ:dk}) remain well-defined.  It is
interesting, then, to see how these quantities behave in a Chern
insulator.  Do each of these quantities remain finite, or does one or
both of them diverge?  Also, what is the behavior of these quantities as
one approaches the NI/CI phase boundary?

\begin{figure}
\begin{center}
\includegraphics[width=\columnwidth]{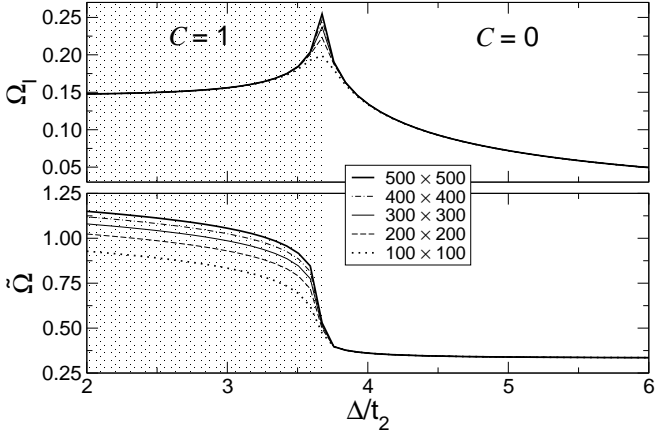}
\end{center}
\caption{\label{fig:omega} Gauge-independent part $\Omega_I$ and
gauge-dependent part $\widetilde{\Omega}$ of the spread functional for
the Haldane model as a function of the $\k$-mesh density.}
\end{figure}

To answer these questions, we have computed the quantities in
Eqs.~(\ref{equ:ik}--\ref{equ:dk}) using the finite-difference versions
of these equations given in Eqs.~(34) and (36) of
Ref.~[\onlinecite{Marzari97}]. For the calculation of the
gauge-dependent part $\widetilde{\Omega}$, we have fixed our gauge such
that $\ket{\psi_\k}$ is real for all $\k$ on the lower-energy site in
the home unit cell. The results are plotted in Fig.~\ref{fig:omega} for
different densities of the $\k$-mesh. It can be seen that $\Omega_I$ is
finite inside both the normal and Chern-insulator regions.  At the
critical value of $\Dcr\approx 3.67$, however, $\Omega_I$ diverges
logarithmically with the number of $\k$-points. Furthermore, 
$\widetilde{\Omega}$ is finite in the normal insulator region, but
diverges logarithmically with the number of $\k$-points for Chern
insulators.  This latter behavior is consistent with the presence of a
vortex in the phases of the $\ket{w_\k}$ around point $\k_a$, which
causes $\A$ to diverge as $|\k-\k_a|^{-1}$ and imparts a logarithmic
divergence to Eq.~(\ref{equ:dk}). It follows that the total spread
$\Omega$ is finite in normal insulators and divergent in Chern
insulators. Heuristically, it is tempting to associate this divergence
with the presence of the metallic chiral edge states that are required
to exist in Chern insulators (see Sec.~\ref{sec:haldane}), but it is
unclear precisely how these features are related. Note that electron
localization in the quantum Hall regime is discussed in detail by R.\
Resta in Ref.~[\onlinecite{Resta05}].

\section{Decay of the density matrix}
\label{sec:densmat}

The decay of the density matrix is a fundamental property of a system
and it is closely connected to the electron localization. It was first
studied by W.\ Kohn for one-dimensional insulators,\cite{Kohn64} and
many others have investigated this topic thereafter.\cite{Resta99,
Souza00, He01, Resta05, Tara02, Tara02_2}

For periodic samples the density matrix is defined as
\begin{equation}\label{equ:rhodef}
\rho(\r,\r') = \frac{\cal A}{(2\pi)^2}\sum_{n=1}^{\text{occ}}\intk\;\psi_{n\k}^*(\r)
    \psi_{n\k}^{}(\r')\;,
\end{equation}
where we assume that the wave functions $\psi_{n\k}$ are normalized to
one unit cell of area $\cal A$. If the wave functions are written in
terms of some basis functions $\phi_{\alpha}^{\k}(\r)$,
\begin{equation}
\psi_{n\k}(\r)=\sum_\alpha C_{n\alpha}^{\k}\phi_{\alpha}^{\k}(\r)\;,
\end{equation}
this becomes
\begin{equation}\label{equ:rhobasis}
\rho(\r,\r') =\frac{\cal A}{(2\pi)^2}
    \sum_{n=1}^{\text{occ}}\sum_{\alpha\beta}\intk\;C_{n\alpha}^{\k*}
    C_{n\beta}^{\k}\,\phi_{\alpha}^{\k*}(\r)\phi_{\beta}^{\k}(\r')\;.
\end{equation}
The $C_{n\alpha}^{\k}$ are the eigenvectors obtained by diagonalizing
the model Hamiltonian---in our case  Eq.~(\ref{equ:hamiltonian}). In a
tight-binding model, the basis functions $\phi_{\alpha}^{\k}(\r)$ are
made up of localized orbitals $\phi$ at sites $\r_\alpha$:
\begin{equation}\label{equ:basisdef}
\phi_{\alpha}^{\k}(\r)=\sum_{\R}
e^{i\k\cdot(\R+\r_{\alpha})}\phi\big(\r-(\R+\r_\alpha)\big)\;.
\end{equation}
Inserting Eq.~(\ref{equ:basisdef}) into Eq.~(\ref{equ:rhobasis}) gives
\begin{equation}\label{equ:rho2}
\rho(\r,\r')=
\sum_{\genfrac{}{}{0pt}{}{\scriptstyle \R\R'}{\scriptstyle \alpha\beta}}
   \xi_{\alpha\beta}(\R'-\R)\,
   \phi^*(\r-\R-\r_\alpha)\,\phi(\r'-\R'-\r_\beta)
\end{equation}
where
\begin{equation}\label{equ:xi}
\xi_{\alpha\beta}(\R) = \frac{\cal A}{(2\pi)^2}
   \sum_{n=1}^{\text{occ}}\intk\;
   C_{n\alpha}^{\k*}C_{n\beta}^{\k}e^{i\k\cdot(\R+\r_{\beta}-
   \r_\alpha)} \;.
\end{equation}
The density matrix cannot be evaluated explicitly without the knowledge
of the orbitals $\phi$, but we can study instead the decay of
$\xi_{\alpha\beta}(\R)$, which essentially has the interpretation of
being a density matrix expressed in a tight-binding representation.

\begin{figure}
\begin{center}
\includegraphics[width=\columnwidth]{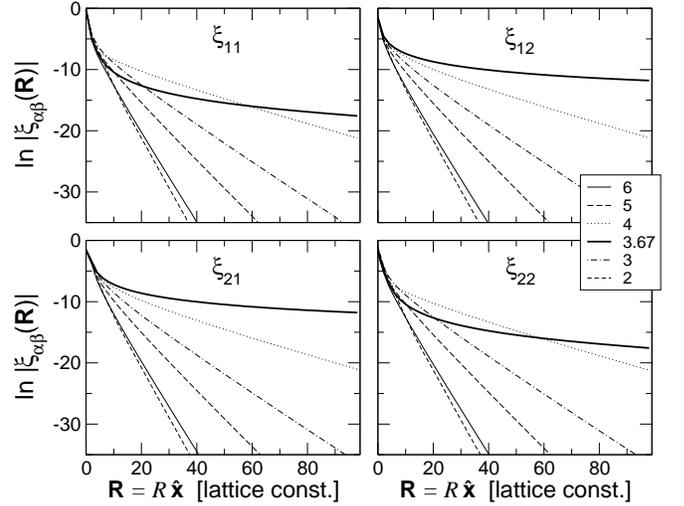}
\end{center}
\caption{\label{fig:densmat} Logarithm of the density-matrix kernel
$|\xi_{\alpha\beta}(R\hat{\x})|$ of the Haldane model for several values
of $\Delta/t_2$.  The linear asymptotic behavior indicates exponential
decay except at $\Dcr\approx3.67$ (solid curves) where the decay is
power-law instead.}
\end{figure}

Calculating the decay of $\xi_{\alpha\beta}(\R)$ in Eq.~(\ref{equ:xi})
numerically is very demanding and the corresponding results are to be
interpreted with caution. To ensure high accuracy, we used a very dense
$\k$-mesh of $2000\times 2000$ points and 128-bit arithmetic. Results
for $\xi_{\alpha\beta}(R\hat{\x})$ (i.e., along the $x$ direction) for
the Haldane model are collected in Fig.~\ref{fig:densmat}. In normal
insulators the density matrix decays exponentially with a power-law
prefactor.\cite{He01} We therefore choose to fit our results according
to $\xi_{\alpha\beta}\sim R^{-a}e^{-bR}$, where $R=|\R|=|R\hat{\x}|$,
and $a$ and $b$ are fit parameters. More specifically, we performed
least-square fits of $\ln|\xi_{\alpha\beta}|$ for distances up to 100
unit cells. For the decay behavior at the NI/CI boundary, we even went
as far as 500 unit cells.

Within fitting error, the best-fit values for the parameter $b$ are the
same for all $\xi_{\alpha\beta}$. Numerical results corresponding to
$\Delta/t_2$ values of 6, 5, 4, 3.67, 3, and 2 are 0.69$\pm$1,
0.43$\pm$1, 0.118$\pm$5, 0.0001$\pm$1, 0.282$\pm$1, and 0.75$\pm$1,
respectively.  In general, when approaching the phase boundary from
either side, the best-fit value of the parameter $b$ decreases and takes
its minimum of zero at \Dcr. In other words, in the normal and
Chern-insulator regions the decay is dominated by the exponential
behavior. However, exactly at the phase boundary the exponential decay
vanishes ($b=0$) and a pure power-law behavior remains, similar to
metals. At \Dcr\ the power-law decay is then characterized by
$a=3.01$$\pm$3 for $\xi_{11}$ and $\xi_{22}$, and $a=2.00$$\pm$2 for
$\xi_{12}$ and $\xi_{21}$, which suggests that the ``true'' values are
the integers 3 and 2. Note that the results depicted in
Fig.~\ref{fig:densmat} correspond to a particular direction in real
space ($\R=R\hat{\x}$). While the decay parameters inside the normal and
Chern-insulator phase depend slightly on the direction, they become
universal at \Dcr. Again, this is a signature of the metallic character.

It is interesting that the power of the power-law decay at the phase
boundary seems to be exactly an integer and that it differs by $1$ for
different  $\xi_{\alpha\beta}$. This behavior can be understood in the
following way: $\xi_{\alpha\beta}(\R)$ of Eq.~(\ref{equ:xi}) is
essentially the Fourier transform of the kernel
$C_{n\alpha}^{\k*}C_{n\beta}^{\k}$ and it is well known that
discontinuities in the kernel determine the decay behavior of the
resulting quantity. In one dimension the discontinuities are related to
the decay like $R^{-(l+1)}$, where $l$ is the number of continuous
derivatives of the kernel.\cite{Goedecker99,Lighthill} Unfortunately, in
two  dimensions the situation is more complex and the resulting BZ
integrals cannot easily be solved analytically. Nevertheless, we give
heuristic  arguments that a similar expression holds for higher
dimensions.

To this end, we solve for analytic expressions of $C_{n\alpha}^{\k}$ by
diagonalizing the Hamiltonian $H(\k)$ in Eq.~(\ref{equ:hamiltonian}). In
turn, we find analytic expressions for the kernel
$C_{n\alpha}^{\k*}C_{n\beta}^{\k}$. Next, we switch to polar coordinates
$\k=(k_x,k_y)\to\k=(k,\phi)$, replace $\Delta/t_2$ by
$\Dcr=3\sqrt{3}\sin\varphi$, and expand the kernel around the Dirac
point in orders of $k$:
\begin{eqnarray}
C_{11}^{\k*}C_{11}^{\k} &=& \frac{1}{2} -
   \frac{3t_2\sin\varphi}{4t_1}k\nn
   &&{}+\frac{5t_2\cos 3\phi\,\sin\varphi}{16\sqrt{3}t_1}k^2\nn
   &&{}+\mathcal{O}(k^3)
\label{equ:oooo}
\\[4mm]
C_{11}^{\k*}C_{12}^{\k} &=& \frac{(1-i\sqrt{3})e^{-i\phi}}{4}\nn
   &&{}+\frac{(3+i\sqrt{3})e^{-i\phi}\sin 3\phi}{48}k\nn
   &&{}+\mathcal{O}(k^2)   
\label{equ:ooot}
\end{eqnarray}

{From} the above expansions it is apparent that
$C_{11}^{\k*}C_{11}^{\k}$ has its first discontinuity in first order in
$k$. Hence, there are $l=1$ continuous derivatives. On the other hand,
due to the $e^{-i\phi}$ term, $C_{11}^{\k*}C_{12}^{\k}$ has already a
discontinuity in zeroth order in $k$, i.e.\ $l=0$. This is consistent
with the numerical results for $\xi_{\alpha\beta}(\R)$ in
Fig.~\ref{fig:densmat} if we assume that the decay in two dimensions is
according to $R^{-(l+2)}$. Equations~(\ref{equ:ooot}) and
(\ref{equ:oooo}) are thus consistent with a decay of $R^{-2}$ and
$R^{-3}$, respectively.

In summary, the numerical and analytical arguments are consistent in
supporting the conclusion that the diagonal and off-diagonal elements of
$\xi_{\alpha\beta}(\R)$ decay as $R^{-3}$ and $R^{-2}$ respectively. An
arbitrary pair of coordinates $\r$ and $\r'$ in Eq.~(\ref{equ:rho2})
will involve a linear combination of contributions coming from diagonal
and off-diagonal terms, so the final conclusion is that the decay of the
density matrix will be as $R^{-2}$ exactly on the NI/CI boundary, and
exponential for any point lying within the normal-insulator or
Chern-insulator phase.

Above, we have evaluated the density matrix $\rho(\r,\r')$ for periodic
samples. For finite samples, we expect a parallel behavior to hold for
points $\r$ and $\r'$ deep inside the bulk.  However, if both points are
chosen to be near the surface of a Chern-insulator sample, one may
expect that the presence of metallic chiral edge states will induce a
power-law decay with the distance between $\r$ and $\r'$ as measured
along the perimeter. Preliminary calculations on finite samples appear
consistent with this picture.

\section{Conclusions}
\label{sec:conclusions}

We have performed numerical and analytical calculations to study the
behavior of several properties of the Haldane model as the system
undergoes a transition from the normal-insulator phase to the
Chern-insulator phase. We first showed how the usual methods of
constructing Wannier functions break down for Chern insulators. We then
investigated several quantities related to electron localization. We
found that the total spread functional, which is finite in normal
insulators, diverges in the case of a Chern insulator. However, when the
spread functional is decomposed into its gauge-independent and
gauge-dependent parts, the former is found to remain finite in a Chern
insulator, while only the latter diverges.  The localization length of
Resta and Sorella,\cite{Resta99} which is related to the
gauge-independent part of the spread functional, thus remains finite. 
However, the localization length increases and diverges logarithmically
as one approaches the NI/CI transition. Similarly, when inspecting the
density matrix, we find that it decays exponentially inside both the
normal and Chern-insulator phases, but that the decay length increases
as the phase boundary is approached, and the behavior crosses over to a
power-law decay exactly at the phase boundary.

We thus find that a system that is sitting right on the NI/CI boundary
has a kind of semimetallic character similar to that of graphene, in
which the valence and conduction bands touch at one (for the Haldane
model) or two (for graphene) Dirac points in the BZ.  When the system is
in the Chern-insulator phase, it still has remnants of metallic behavior
in the presence of metallic edge states, the divergence of the total
spread functional, and the difficulty of constructing Wannier functions.

Our results were obtained here for a specific realization of a Chern
insulator, namely, the Haldane model.  While it seems very likely that
the localization properties found here will apply to other
Chern-insulator systems, it remains to test this hypothesis by carrying
out similar studies on other systems. It would also be of considerable
interest to extend the current study to three-dimensional
Chern-insulator crystals, and to continuum, as opposed to tight-binding,
models. These could be fruitful avenues for future investigations.

\begin{acknowledgments}
We acknowledge fruitful discussions with R.~Resta and D.~Ceresoli. This
work was supported by NSF grants No.\ DMR-0233925 and DMR-0549198.
\end{acknowledgments}



\begin{thebibliography}{10}
\bibitem{Explan-ni} For an interacting system with ground-state
   degeneracy, the Chern invariant is rationally related to
   a $\G$-vector.
\end{thebibliography}
\end{document}